# Low SNR Capacity of Fading Channels with Peak and Average Power Constraints

Vignesh Sethuraman, *Member, IEEE*, and Bruce Hajek, *Fellow, IEEE*

### Abstract

Flat-fading channels that are correlated in time are considered under peak and average power constraints. For discrete-time channels, a new upper bound on the capacity per unit time is derived. A low SNR analysis of a full-scattering vector channel is used to derive a complimentary lower bound. Together, these bounds allow us to identify the exact scaling of channel capacity for a fixed peak to average ratio, as the average power converges to zero. The upper bound is also asymptotically tight as the average power converges to zero for a fixed peak power.

For a continuous time infinite bandwidth channel, Viterbi identified the capacity for M-FSK modulation. Recently, Zhang and Laneman showed that the capacity can be achieved with non-bursty signaling (QPSK). An additional contribution of this paper is to obtain similar results under peak and average power constraints.

## I. INTRODUCTION

A Rayleigh flat-fading channel with correlation in time is considered. There is no channel state information at either the transmitter or the receiver. The fading process is modeled as a stationary ergodic process. An average power constraint is imposed at the transmitter. In the low SNR regime, it is known that capacity achieving input signals are bursty [1, 2]. To limit this behavior, a hard peak constraint, in addition to the average power constraint, is considered. The focus of this paper is the channel capacity with peak and average power constraints in the low SNR regime. For similar work in the high SNR regime, see [3] and references there-in.

The capacity of this channel in the low SNR regime has recently been of heightened research interest [4–6]. In these works, a lower bound is obtained through constructive methods, and is compared to an upper bound based on capacity per unit energy [7, 8]. While the bounds are close for channels with memory, an asymptotic analysis at low SNR reveals a gap between these bounds.

The main result in this work is the complete identification of the asymptotic behavior of the capacity of this channel at low SNR. To this end, a new upper bound on the channel capacity is derived. To obtain a matching lower bound, a full-scattering vector channel is considered. For a surprisingly simple choice of input distribution, the mutual information rate of this channel, when specialized to flat fading, is shown to coincide with the upper bound. Incidentally, it is also established in this work that the bounds in [4–6] are not, in general, asymptotically tight at low SNR. Nevertheless, when there is only a peak constraint, a lower bound in [4, 6] is found to be asymptotically tight when there is sufficient memory (correlation in time) in the channel. Interestingly, an upper bound derived in [3] for tightness at high SNR is found to be tight at low SNR too.

[1]This work was supported in part by the National Science Foundation Grant NSF ITR 00-85929, and the Motorola Center for Communication Graduate Fellowship.

A continuous time version of the flat fading channel is considered.[2] The channel capacity for M-FSK modulation is calculated in [9]. The input signaling here is bursty in frequency. Zhang and Laneman [4] find the capacity of this channel under a peak constraint, by showing that the rate achieved using a specific constructive scheme matches an upper bound obtained using capacity per unit energy. Also, since the constructive scheme can use M-PSK, it is not bursty in either time or frequency. Here, similar results are established under both peak and average power constraints.

The paper is organized as follows. In Section II, the channel model is defined and some background on the channel capacity is developed. Upper bounds are presented in Section III. The main result, namely, the asymptotic behavior of capacity at vanishing peak and average powers, is stated and proved in Section IV. The material on continuous time channels is presented in Section V. Interesting directions of future research are outlined in Section VI.

## II. Preliminaries

Consider a single-user discrete-time channel without channel state information at either transmitter or receiver. The channel includes additive noise and multiplicative noise (Rayleigh flat-fading) and is specified by

$$Y_k = \sqrt{\rho} Z_k H_k + W_k; \ k \in \mathbb{Z} \tag{1}$$

where $X_k = \sqrt{\rho} Z_k$ is the input, $H$ is the fading process, $W$ is an additive noise process, and $Y$ is the output. The fading and the additive noise processes are mutually independent, and jointly independent of the input. The additive noise $W$ is modeled as an i.i.d. proper complex normal (PCN) process with unit variance. The fading process $H$ is also PCN, but is correlated in time with an autocorrelation function $\{R_H(k) : k \in \mathbb{Z}\}$. Further, $H$ is assumed to be ergodic. A peak power constraint $|X_i|^2 \leq \rho$ for some $\rho > 0$ is imposed at the transmitter. This translates to

$$|Z_i|^2 \leq 1 \tag{2}$$

An average power constraint, specified through the peak to average ratio $\beta \geq 1$ and the peak power constraint $\rho$, is also imposed at the transmitter.

$$E[|Z_i|^2] \leq \frac{1}{\beta} \tag{3}$$

In [7], the information-theoretic capacity per unit time for this channel, in the presence of the above defined peak and average power constraints on the input, is defined and shown to be equal to the operational capacity. The capacity per unit time of this channel is denoted, in this manuscript, by $C(\rho, \beta)$.

Consider the following model of a full-scattering vector channel.

$$Y_{1 \times n} = \sqrt{\rho} Z_{1 \times n} \widehat{H}_{n \times n} + W_{1 \times n} \tag{4}$$

A deterministic signal $X_{1 \times n} = \sqrt{\rho} Z_{1 \times n}$ is the input and $Y$ is the observed output. Here, the random variables $\widehat{H}_{i,j}$ are elements of the fading process. They are modeled as PCN with unit variance and are allowed to be correlated with each other, but jointly independent of the additive noise term $W$, which can be thought of as part of an i.i.d. PCN process with unit variance. Peak (2) and average power (3) constraints can be applied to this model too. Let $\mu$ be a distribution on $Z_{1 \times n}$ satisfying the peak constraint (2). For low values of peak constraint $\rho$, the mutual information between the output and the input is well approximated by the leading terms of the Taylor's series expansion.

---

[2] A flat fading channel model in continuous-time comes with the caveat that such a model is questionable when input symbol durations are smaller than the delay spread of the fading process. Still, this model isn't bad for a first-cut approximation of continuous-time channels.

*Lemma 2.1:* For small values of $\rho$, $I(Z; Y)$ is given by

$$I(Z; Y) = \frac{\rho^2}{2} \left( E_{Z \sim \mu} \left[ Tr(K_Z^2) \right] - Tr(K_\mu^2) \right) + o(\rho^2) \tag{5}$$

where

$$K_Z = E_{\widehat{H}} [\widehat{H}^\dagger Z^\dagger Z \widehat{H}] \tag{6}$$

and

$$K_\mu = E_{Z \sim \mu} [K_Z]. \tag{7}$$

Here, $o(x)$ is used in the sense that $\lim_{x \to 0} \frac{o(x)}{x} = 0$. The proof of above lemma is given in Appendix I.

Henceforth, in this manuscript, $\widehat{H}$ is assumed to be diagonal with entries $\widehat{H}_{i,i} = H_i$. Then, (4) is simply a vector notation of the first $n$ channel uses of (1).

It is of interest to calculate the maximum mutual information rate

$$C_n(\rho, \beta) = \frac{1}{n} \sup_{P_{X_1^n} : \rho, \beta} I(X_1^n; Y_1^n) \tag{8}$$

that can be achieved in (1) with the constraints (2), (3). Here, $X_1^n$ refers to the set $\{X_i : 1 \leq i \leq n\}$, and $\{P_{X_1^n} : \rho, \beta\}$ is the set of distributions on $X_1^n$ that satisfy a peak power constraint $\rho$, and peak to average ratio $\beta$. Since the information-theoretic capacity exists [7],

$$C(\rho, \beta) = \lim_{n \to \infty} C_n(\rho, \beta) \tag{9}$$

## III. UPPER BOUNDS

Two upper bounds on the channel capacity are presented. Let

$$U(\rho, \beta) = \log \left( 1 + \rho \theta(\rho, \beta) \right) - \theta(\rho, \beta) I(\rho) \tag{10}$$

where

$$I(\rho) = \int_0^{2\pi} \log(1 + \rho S_H(\omega)) \, \frac{d\omega}{2\pi} \tag{11}$$

$$\theta(\rho, \beta) = \min \left( \frac{1}{\beta}, \frac{1}{I(\rho)} - \frac{1}{\rho} \right) \tag{12}$$

*Proposition 3.1:* For any peak constraint $\rho > 0$ and peak to average ratio $\beta \geq 1$,

$$C(\rho, \beta) \leq U(\rho, \beta) \tag{13}$$

*Proof:* We shall first establish that, for any $n \in \mathbb{N}$,

$$C_n(\rho, \beta) \leq U(\rho, \beta). \tag{14}$$

The proposition then follows from (9) and (14).

Consider $C_n(\rho, \beta)$ for some $n > 0$.

$$C_n(\rho, \beta) = \sup_{P_X : \rho, \beta} \frac{1}{n} I(X_1^n; Y_1^n) \tag{15}$$

$$= \sup_{P_X : \rho, \beta} \frac{1}{n} \sum_{i=1}^n I(Y_i; X_1^n | Y_1^{i-1}) \tag{16}$$

by the chain rule of mutual information. Denoting the set $\{X_i : |i| < \infty\}$ by $X$, it can be shown that

$$I(Y_i; X_1^n | Y_1^{i-1}) = I(Y_i; X | Y_1^{i-1}). \tag{17}$$

for $1 \leq i \leq n$. This follows from a generalization of the fact that, conditioned on the current input and past inputs and outputs, the current output is independent of future inputs due to causality of the channel.

*Lemma 3.1:* For any $i \in \mathbb{N}$, the mutual information $I(Y_i; X | Y_1^{i-1})$ has the following upper bound.

$$I(Y_i; X | Y_1^{i-1}) \leq \log(1 + \rho\theta) - \theta I(\rho) \tag{18}$$

where $\theta = E[|Z_i|^2]$.

The proof of the above lemma is given in Appendix II. Since the above lemma holds for all input distributions satisfying the power constraints, the following upper bound on $C_n(\rho, \beta)$ is obtained.

$$
\begin{aligned}
C_n(\rho, \beta) &= \sup_{P_X : \ \rho, \beta} I(Y_n; X | Y_F) \tag{19} \\
&\leq \max_{0 \leq \theta \leq \frac{1}{\beta}} \log(1 + \rho\theta) - \theta I(\rho) \tag{20}
\end{aligned}
$$

The proof of (14) is completed by noting that the maximum in the above expression is attained at

$$\theta = \min(\frac{1}{\beta}, \frac{1}{I(\rho)} - \frac{1}{\rho}). \tag{21}$$

∎

A second upper bound on the capacity per unit time is adapted from [3]:

$$C(\rho, \beta) \leq U_{pred}(\rho, \beta) \tag{22}$$

where

$$U_{pred}(\rho, \beta) = \max_{P_{ave} \leq \rho/\beta} \left\{ \sup_{P_{X_0} : \ \rho, P_{ave}} I(X_0; Y_0) + \log\left(\frac{1 + P_{ave}}{1 + P_{ave}\frac{e^{I(\rho)} - 1}{\rho}}\right) \right\} \tag{23}$$

It is noted that, while the initial part of the derivation of $U(\rho, \beta)$ is similar to that of $U_{pred}(\rho, \beta)$ in [3], the evaluation of $U(\rho, \beta)$ does not require an optimization over input distributions. The following lower bound is quoted from [6].

$$C_{l1}(\rho, \beta) = \frac{1}{\beta} I(X_0; Y_0 | X_{-\infty}^{-1}, Y_{-\infty}^{-1}). \tag{24}$$

where the input $X$ is an i.i.d. process such that, for each $i$, $X_i$ is constant amplitude $\sqrt{\rho}$ and zero mean. The following upper bound, quoted from [6, (40)] is derived from the analysis of capacity per unit energy.

$$C_u(\rho, \beta) = \frac{1}{\beta}(\rho - I(\rho)) \tag{25}$$

## IV. Capacity asymptote at small peak and average powers

In this section, a low SNR asymptotic analysis is performed as follows: the peak to average ratio, $\beta$, is fixed, and vanishingly small values of the peak power constraint $\rho$ (and, consequently, of the average power constraint) are considered. The asymptotic behavior of the capacity per unit time $C(\rho, \beta)$ is captured by the following result.

*Proposition 4.1:* For $\beta > 0$, $\lim_{\rho \to 0} C(\rho, \beta)/\rho^2$ exists and is given by

$$\lim_{\rho \to 0} \frac{C(\rho, \beta)}{\rho^2} = \begin{cases} \frac{\lambda_\infty^2}{8} & \text{if } \frac{\lambda_\infty}{2} \leq \frac{1}{\beta} \\ \frac{\lambda_\infty}{2\beta} - \frac{1}{2\beta^2} & \text{if } \frac{\lambda_\infty}{2} \geq \frac{1}{\beta} \end{cases} \tag{26}$$

where $\lambda_\infty$ is given by

$$\lambda_\infty = \sum_{i=-\infty}^{\infty} |R_H(i)|^2 \tag{27}$$

$$= \int_0^{2\pi} |S_H(\omega)|^2 \, \frac{d\omega}{2\pi} \tag{28}$$

Here, $S_H$ is the power spectral density (if it exists) of the fading process $H$. (It is assumed here that the integral in (28) exists and is finite.)

*Proof:* Let the expression on the RHS of (26) be denoted by $f(\beta)$. The following lemma is useful in proving Proposition 4.1.

*Lemma 4.1:*

$$\lim_{\rho \to 0} \frac{U(\rho, \beta)}{\rho^2} = f(\beta) \tag{29}$$

The proof of the above lemma is in Appendix III. From Lemma 4.1 and Proposition 3.1, it follows that

$$\limsup_{\rho \to 0} \frac{C(\rho, \beta)}{\rho^2} \le f(\beta) \tag{30}$$

The following lemma completes the proof of Proposition 4.1.

*Lemma 4.2:*

$$\liminf_{\rho \to 0} \frac{C(\rho, \beta)}{\rho^2} \ge f(\beta) \tag{31}$$

The proof of Lemma 4.2 is given in Appendix IV. ∎

Proposition 4.1 and Lemma 4.1 together prove that $U(\rho, \beta)$ is asymptotically tight as peak power $\rho \to 0$ for a fixed $\beta > 0$. The second upper bound $U_{pred}(\rho, \beta)$ is also found to be asymptotically tight.

It can be shown that, for a fixed $\beta$, the bounds in (24) and (25) have the following limits.

$$\lim_{\rho \to 0} \frac{C_{l1}(\rho, \beta)}{\rho^2} = \frac{\lambda_\infty - 1}{2\beta} \tag{32}$$

$$\lim_{\rho \to 0} \frac{C_u(\rho, \beta)}{\rho^2} = \frac{\lambda_\infty}{2\beta} \tag{33}$$

The above limits were derived in [4] for the specific case when $\beta = 1$. By comparing the above limits with $f(\beta)$, it is seen that the bounds in [4, 6] are not, in general, asymptotically tight. Nevertheless, when there is only a peak constraint ($\beta = 1$), the lower bound $C_{l1}(\rho, \beta)$ is tight when $\lambda_\infty \ge 2$; i.e. when there is sufficient memory in the fading process.

## V. Continuous time channels

A continuous-time version of the channel modeled in (1) is considered. Following [7, (20)], the channel model is given by:

$$Y(t) = H(t)X(t) + W(t), \ 0 \le t \le T \tag{34}$$

Here, $W(t)$ is a complex proper Gaussian white noise process with $E[W(s)\overline{W(t)}] = \delta(s-t)$, and $(H(t) : -\infty < t < \infty)$ is a stationary ergodic proper complex Gaussian process such that $E[|H(t)|^2] = 1$. A deterministic signal $X = (X(t) : 0 \le t \le T)$, where $T$ is the duration of the signal, is the channel input and $Y(t)$ is the observed output.

As in [7], both average power and peak power constraints are imposed on the transmitter. The power constraints are defined as follows:

$$\frac{1}{T} \int_0^T |X(t)|^2 dt \quad \leq \quad P_{ave}, \tag{35}$$

$$\sup_{0 \leq t \leq T} |X(t)|^2 \quad \leq \quad P_{peak}. \tag{36}$$

There is no bandwidth constraint, other than that the input codewords are required to be Borel measurable functions of $t$.

*Proposition 5.1:* The capacity per unit time of the continuous-time channel under power constraints $P_{ave}$ and $P_{peak}$ is given by

$$C(P_{ave}, P_{peak}) = P_{ave} - \frac{P_{ave}}{P_{peak}} I(P_{peak}) \tag{37}$$

where

$$I(P_{peak}) = \int_{-\infty}^{\infty} \log(1 + P_{peak} S_H(\omega)) \frac{d\omega}{2\pi}. \tag{38}$$

where $S_H(\omega)$ denotes the density of the absolutely continuous component of the power spectral measure of $H$.

An outline of the proof is now provided.

Following [7, (12) and Prop. 3.3], the capacity per unit energy $C_p(P_{peak})$ of this channel gives the following upper bound on $C(P_{ave}, P_{peak})$.

$$C(P_{ave}, P_{peak}) \leq P_{ave} - \frac{P_{ave}}{P_{peak}} I(P_{peak}) \tag{39}$$

To obtain a matching lower bound, firstly, a discrete-time channel is derived from the continuous-time channel by constraining the continuous-time transceiver to using only certain encoding and decoding methods. An information theoretic lower bound on the capacity of the discrete-time channel, presented in [6], is then used. Next, the connection between mutual information and the mean square error in estimating the input given the output [10] is used to evaluate the lower bound. This lower bound is then translated to continuous-time and is shown to coincide with the upper bound, thus giving the capacity per unit time in closed form.

Zhang and Laneman [4] observe that the upper bound (39) is tight when $P_{ave} = P_{peak}$ by obtaining a matching lower bound using a constructive scheme. The scheme uses an interleaver to convert the channel with correlated fading into a set of parallel sub-channels (PSC), with i.i.d. fading in each PSC. While this works for channels with finite memory, i.e. there exists $K > 0$ such that $R_H(k) = 0$ for all $|k| > K$, it is unclear how the proof extends to more general ergodic fading channels with infinite memory. In our proof technique, this problem is avoided by obtaining a lower bound using purely information-theoretic methods (see [6]).

## VI. FUTURE DIRECTIONS

The low SNR regime is largely motivated by broadband channels, where the total transmit power is constrained while the available number of degrees of freedom is virtually unlimited. Since broadband channels are, in truth, multipath channels, a specular multipath fading process models such channels in a more realistic manner. We seek to investigate the asymptotic capacity of such multipath fading channels. The vector channel model (4) and Lemma 2.1 apply to these channels.

Another direction is to obtain capacity bounds that are good for all SNR. The bound $U_{pred}$ (23) is an example of such a bound. This bound is asymptotically tight at both high SNR [3] and at low SNR (for $\beta \geq 2$). Deriving and comparing matching lower bounds to such upper bounds will help in characterizing the capacity of correlated fading channels for all values of SNR.

APPENDIX I

PROOF OF LEMMA 2.1

Consider the output of the channel modeled in (1). Then, $Y^\dagger Y_{n \times n}$ is given by

$$Y^\dagger Y_{n \times n} = \rho H^\dagger Z^\dagger Z H + W^\dagger W \tag{40}$$

Denoting $cov(Y|Z)$ by $K_Y$,

$$K_Y = \rho K_Z + I \tag{41}$$

where $K_Z$ is defined in (6).

The probability density function (pdf) of $Y$ conditioned on $Z$ is given by

$$q(Y|Z) = \frac{\exp(-Y K_Y^{-1} Y^\dagger)}{\pi^n \det(K_Y)} \tag{42}$$

So,

$$\frac{q(Y|Z)}{q(Y|0)} = \frac{\exp(-Y K_Y^{-1} Y^\dagger + Y Y^\dagger)}{\det(K_Y)} \tag{43}$$

Following [11], let

$$\Delta q(Y|Z) = \frac{q(Y|Z)}{q(Y|0)} - 1 \tag{44}$$

Since, for small $\rho$,

$$K_Y^{-1} = I - \rho K_Z + o(\rho) \tag{45}$$

and

$$\det(K_Y) = 1 + \rho Tr(K_Z) + o(\rho) \tag{46}$$

we have,

$$\Delta q(Y|Z) = \frac{\exp(\rho Y K_Z Y^\dagger + o(\rho))}{1 + \rho Tr(K_Z) + o(\rho)} - 1 \tag{47}$$

$$= \rho \left( Y K_Z Y^\dagger - Tr(K_Z) \right) + o(\rho) \tag{48}$$

Here, $o(x)$ is used in the sense that $\lim_{x \to 0} \frac{o(x)}{x} = 0$. Using [11, (13),(14)] and proceeding as in [11, §VII], it can be shown that

$$\widetilde{h}(Y|Z) = -E_{X \sim \mu_\epsilon} \left[ E_0 \left[ \frac{q(Y|Z)}{q(Y|0)} \log \left( \frac{q(Y|Z)}{q(Y|0)} \right) \right] \right] \tag{49}$$

$$= -\frac{\rho^2}{2} E_{Z \sim \mu} \left[ E_0 \left[ \left( Y K_Z Y^\dagger - Tr(K_Z) \right)^2 \right] \right] + o(\rho^2) \tag{50}$$

$$\widetilde{h}(Y) = -E_0 \left[ \frac{q(Y|\mu_\epsilon)}{q(Y|0)} \log \left( \frac{q(Y|\mu_\epsilon)}{q(Y|0)} \right) \right] \tag{51}$$

$$= -\frac{\rho^2}{2} E_0 \left[ \left( E_{Z \sim \mu} \left[ Y K_Z Y^\dagger - Tr(K_Z) \right] \right)^2 \right] + o(\rho^2) \tag{52}$$

Further,

$$E_0 \left[ Y K_Z Y^\dagger \right] = Tr(K_Z) \tag{53}$$

$$E_0 \left[ \left( Y K_Z Y^\dagger \right)^2 \right] = (Tr(K_Z))^2 + Tr(K_Z^2) \tag{54}$$

This, when substituted in (50) yields

$$\widetilde{h}(Y|Z) = -\frac{\rho^2}{2} E_{Z \sim \mu} \left[ Tr(K_Z^2) \right] + o(\rho^2) \tag{55}$$

It can be shown in a similar fashion that

$$\widetilde{h}(Y) = -\frac{\rho^2}{2} Tr(K_\mu^2) + o(\rho^2) \tag{56}$$

where $K_\mu$ is defined in (7). Since the mutual information $I[Z; Y)$ is given by

$$I(Z; Y) = \widetilde{h}(Y) - \widetilde{h}(Y|Z), \tag{57}$$

for low SNR, $I(Z; Y)$ is given by

$$I(Z; Y) = \frac{\rho^2}{2} \left( E_{Z \sim \mu} \left[ Tr(K_Z^2) \right] - Tr(K_\mu^2) \right) + o(\rho^2) \tag{58}$$

## Appendix II

## Proof of Lemma 3.1

First, we shall prove the following lemma.

*Lemma 2.1:* For any $i \in \mathbb{Z}$ and $F \subset \mathbb{Z} - \{i\}$, the mutual information $I(Y_i; X|Y_F)$ has the following upper bound.

$$I(Y_i; X|Y_F) \leq \log(1 + \rho\theta) - \theta \log(1 + \rho\bar{\sigma}_{i|F}^2)] \tag{59}$$

where $\theta = E[|Z_i|^2]$.

*Proof:*

$$I(Y_i; X|Y_F) = I(Y_i; X, Y_F) - I(Y_i; Y_F) \tag{60}$$
$$\leq I(Y_i; X, Y_F) \tag{61}$$

The output $Y_i$ can be expressed as follows.

$$Y_i = H_i X_i + W_i \tag{62}$$
$$= (\widehat{H}_i + \widetilde{H}_i) X_i + W_i \tag{63}$$

where $\widehat{H}_i$ is the following MMSE estimate of $H_i$:

$$\widehat{H}_i = E[H_i|X_F, Y_F] \tag{64}$$

and, $\widetilde{H}_i$ is the error in estimation.

$$\widetilde{H}_i = H_i - \widehat{H}_i \tag{65}$$

Let the variance of the error be denoted by

$$\sigma_{i|F}^2 = E[|\widetilde{H}_i|^2] \tag{66}$$

It is noted that $\sigma_{i|F}^2$ is a function of $X_F$. Let $\bar{\sigma}_{i|F}^2$ be the minimum of $\sigma_{i|F}^2(X_F)$ over all $X$ satisfying the peak constraint. This minimum is achieved when $X_k = \rho$ for all $k \in F$. This value of $\bar{\sigma}_{i|F}^2$ is a constant independent of the distribution of $X$ (see [12, pg. 440]).

It follows from (63) that, given $X$ and $Y_F$, $Y_i$ is PCN with variance $1 + |X_i|^2 \sigma_{i|F}^2$. So,

$$h(Y_i|X, Y_F) = \log(\pi e) + E[\log(1 + |X_i|^2 \sigma_{i|F}^2)] \tag{67}$$
$$\geq \log(\pi e) + E[\log(1 + |X_i|^2 \bar{\sigma}_{i|F}^2)] \tag{68}$$

Also, the (unconditional) variance of $Y_i$ is $1 + E[|X_i|^2]$. So,

$$h(Y_i) \leq \log(\pi e) + \log(1 + E[|X_i|^2]) \tag{69}$$

So,

$$
\begin{align}
I(Y_i; X|Y_F) &\leq \log(1 + E[|X_i|^2]) - E[\log(1 + |X_i|^2 \bar{\sigma}_{i|F}^2)] \tag{70} \\
&= \log(1 + \rho E[|Z_i|^2]) - E[\log(1 + \rho|Z_i|^2 \bar{\sigma}_{i|F}^2)] \tag{71}
\end{align}
$$

If $E[|Z_i|^2] = \theta$, then $0 \leq \theta \leq \frac{1}{\beta}$ from the average power constraint, and due to the concavity of log, the upper bound is maximized by

$$
|Z_i|^2 = \begin{cases} 1 & \text{with probability} \quad \theta \\ 0 & \text{with probability} \quad 1 - \theta \end{cases} \tag{72}
$$

This yields the following lemma. ∎

It is noted that, if $F \subset (-\infty, n-1]$, then

$$
\bar{\sigma}_{n|F}^2 \leq \bar{\sigma}_{n|(-\infty, n-1]}^2 \tag{73}
$$

The causal prediction error $\bar{\sigma}_{n|(-\infty, n-1]}^2$ is clearly independent of $n$, and can be equivalently denoted by $\bar{\sigma}_{0|(-\infty, -1]}^2$. Following standard results in estimation theory, the causal prediction error is given by

$$
\sigma_{0|(-\infty, -1]}^2 = \frac{1}{\rho} \left( \exp(I(\rho)) - 1 \right) \tag{74}
$$

where $I(\rho)$ is given by eq:Irho. From Lemma 2.1, it follows that

$$
\begin{align}
I(Y_n; X|Y_F) &\leq \log(1 + \rho\theta) - \theta \log(1 + \rho\sigma_{0|(-\infty, -1]}^2) \tag{75} \\
&= \log(1 + \rho\theta) - \theta I(\rho) \tag{76}
\end{align}
$$

where $\theta = E[|Z_i|^2]$. This completes the proof of Lemma 3.1.

For the interested reader, Lemma 2.1 can be applied to any $F \subset \mathbb{Z} - \{n\}$ in the following manner.

$$
\begin{align}
\bar{\sigma}_{n|F}^2 &\leq \bar{\sigma}_{n|\mathbb{Z}-\{n\}}^2 \tag{77} \\
&= 1 - \rho \int_0^{2\pi} \frac{S_H^2(\omega)}{1 + \rho S_H(\omega)} \frac{d\omega}{2\pi} \tag{78}
\end{align}
$$

## Appendix III
## Proof of Lemma 4.1

Lemma 4.1 is proved as follows.

First, the asymptotic behaviour of the upper bound $U(\rho, \beta)$ is analyzed. For small values of $\rho$, the expression $I(\rho)$, as given in (11), can be approximated as follows.

$$
\begin{align}
I(\rho) &= \int_0^{2\pi} \log(1 + \rho S_H(\omega)) \frac{d\omega}{2\pi} \tag{79} \\
&= \int_0^{2\pi} \left( \rho S_H(\omega) - \frac{\rho^2}{2} S_H^2(\omega) + \frac{\rho^3}{3} S_H^3(\omega) \right) \frac{d\omega}{2\pi} + o(\rho^3) \tag{80} \\
&= \rho - \frac{\rho^2}{2} \lambda_\infty + o(\rho^2) \tag{81}
\end{align}
$$

where $\lambda_\infty$ is given by (28). Using the above Taylor's series expansion of $I(\rho)$, it is easy to show that

$$
\lim_{\rho \to 0} \frac{1}{I(\rho)} - \frac{1}{\rho} = \frac{\lambda_\infty}{2} \tag{82}
$$

The asymptotic behaviour of $U(\rho, \beta)$ as a function of $\rho$ depends on whether $\frac{1}{\beta} > \frac{\lambda_\infty}{2}$. This is considered in the following two cases.

If $\frac{1}{\beta} < \frac{\lambda_\infty}{2}$, then there exists a $\rho_0 > 0$ small enough such that, for all $\rho < \rho_0$,

$$\frac{1}{\beta} \le \frac{1}{I(\rho)} - \frac{1}{\rho}. \tag{83}$$

For such $\rho$, $U(\rho, \beta)$ is given by

$$U(\rho, \beta) = \log(1 + \frac{\rho}{\beta}) - \frac{1}{\beta} I(\rho) \tag{84}$$

Then, for fixed $\beta$, the asymptotic behaviour of $U(\rho, \beta)$ for small values of $\rho$ is given by

$$\lim_{\rho \to 0} \frac{U(\rho, \beta)}{\rho^2} = \lim_{\rho \to 0} \frac{1}{\rho^2} \left( \log(1 + \frac{\rho}{\beta}) - \frac{1}{\beta} I(\rho) \right) \tag{85}$$

$$= \frac{1}{2\beta} \left( \lambda_\infty - \frac{1}{\beta} \right) \tag{86}$$

Similarly, when $\frac{1}{\beta} > \frac{\lambda_\infty}{2}$, then there exists a $\rho_0 > 0$ such that , for all $\rho < \rho_0$,

$$\frac{1}{\beta} \ge \frac{1}{I(\rho)} - \frac{1}{\rho}. \tag{87}$$

Then, $U(\rho, \beta)$ is given by

$$U(\rho, \beta) = -\log \frac{I(\rho)}{\rho} - 1 + \frac{I(\rho)}{\rho} \tag{88}$$

$$= \left( \frac{I(\rho)}{\rho} - 1 \right) - \log \left\{ 1 + \left( \frac{I(\rho)}{\rho} - 1 \right) \right\} \tag{89}$$

For small values of $\rho$, $\frac{I(\rho)}{\rho} - 1$ behaves as follows.

$$\frac{I(\rho)}{\rho} - 1 = -\rho \frac{\lambda_\infty}{2} - \rho^2 \frac{\nu_\infty}{3} + o(\rho^2) \tag{90}$$

From (89) and (90), and using the fact that, for $0 < x \ll 1$,

$$x - \log(1 + x) = x^2/2 + o(x^2), \tag{91}$$

the asymptotic behaviour of $U(\rho, \beta)$ in the limit $\rho \to 0$ is given by

$$\lim_{\rho \to 0} \frac{U(\rho, \beta)}{\rho^2} = \lim_{\rho \to 0} \frac{1}{2\rho^2} \left( \frac{I(\rho)}{\rho} - 1 \right)^2 \tag{92}$$

$$= \frac{\lambda_\infty^2}{8} \tag{93}$$

From (86) and (93), it follows that

$$\lim_{\rho \to 0} \frac{U(\rho, \beta)}{\rho^2} = f(\beta) \tag{94}$$

## Appendix IV
## Proof of Lemma 4.2

Setting $Z = (Z_1, \ldots, Z_n)$ and $V_i = |Z_i|^2$ in Lemma 2.1 (where $\widehat{H}$ is diagonal and $\widehat{H}_{i,i} = H_i$), we have

$$Tr(K_Z^2) = \sum_{i=1}^n |V_i|^2 + 2 \sum_{i=1}^n \sum_{j>i}^n |R_H(i-j)|^2 V_i V_j \tag{95}$$

$$Tr(K_\mu^2) = \sum_{i=1}^n E^2[V_i] + 2 \sum_{i=1}^n \sum_{j>i}^n |R_H(i-j)|^2 |E[Z_i^* Z_j]|^2 \tag{96}$$

Substituting the above in (5),

$$
\begin{aligned}
I(Z;Y) \;=\;& \left\{ \sum_{i=1}^{n} E_{Z \sim \mu}\left[ |V_i|^2 \right] + 2 \sum_{i=1}^{n} \sum_{j>i}^{n} |R_H(i-j)|^2 E_{Z \sim \mu}\left[ V_i V_j \right] \right. \\
& \left. - \sum_{i=1}^{n} E_{Z \sim \mu}^2[V_i] - 2 \sum_{i=1}^{n} \sum_{j>i}^{n} |R_H(i-j)|^2 \, |E_{Z \sim \mu}[Z_i^* Z_j]|^2 \right\} \frac{\rho^2}{2} + o(\rho^2)
\end{aligned}
\tag{97}
$$

Consider the following distribution, $\widehat{\mu}$, on $Z_1^n$. Let

$$
|Z_i| = I_{\{U \le a\}}
\tag{98}
$$

where $U$ is an auxiliary random variable uniformly distributed in $[0,1]$, and $a \in [0, 1/\beta]$ (to satisfy the average power constraint.) Further, let the phases of $Z_i$ be independent and uniformly distributed in $[0, 2\pi]$, for all $i$. So, for each $i$, $|Z_i| \in \{0, 1\}$. Further, for any $i$, $j$,

$$
E[V_i] = E[V_i V_j] = a
\tag{99}
$$

Also, $E[Z_i^* Z_j] = 0$ if $i \ne j$. The corresponding mutual information rate is obtained by substituting the above in (97).

$$
\frac{1}{n} I(Z_1^n; Y_1^n) = \frac{1}{2}\left( a\lambda_n - a^2 \right) + o(\rho^2)
\tag{100}
$$

where

$$
\lambda_n = \frac{1}{n} \sum_{i=1}^{n} \sum_{j=1}^{n} |R_H(i-j)|^2
\tag{101}
$$

Set

$$
a = \begin{cases} \frac{\lambda_n}{2} & \text{if } \frac{\lambda_n}{2} \le \frac{1}{\beta} \\ \frac{1}{\beta} & \text{else} \end{cases}
\tag{102}
$$

It can be checked that, for the above choice of $a$, the distribution $\widehat{\mu}$ does not violate the power constraints. Let the corresponding mutual information rate be denoted by $L_n(\rho, \beta)$. Then, $L_n(\rho, \beta)$ is given by:

$$
L_n(\rho, \beta) = \begin{cases} \frac{\lambda_n^2}{8} \rho^2 + o(\rho^2) & \text{if } \frac{\lambda_n}{2} \le \frac{1}{\beta} \\ \left( \frac{\lambda_n}{2\beta} - \frac{1}{2\beta^2} \right) \rho^2 + o(\rho^2) & \text{if } \frac{\lambda_n}{2} \ge \frac{1}{\beta} \end{cases}
\tag{103}
$$

From (8) and (9), it follows that

$$
C(\rho, \beta) \ge L_n(\rho, \beta)
\tag{104}
$$

Dividing by $\rho^2$ and taking the limit $\rho \to 0$,

$$
\liminf_{\rho \to 0} \frac{C(\rho, \beta)}{\rho^2} \;\ge\; \lim_{\rho \to 0} \frac{L_n(\rho, \beta)}{\rho^2}
\tag{105}
$$

$$
=\; \begin{cases} \frac{\lambda_n^2}{8} & \text{if } \frac{\lambda_n}{2} \le \frac{1}{\beta} \\ \left( \frac{\lambda_n}{2\beta} - \frac{1}{2\beta^2} \right) & \text{if } \frac{\lambda_n}{2} \ge \frac{1}{\beta} \end{cases}
\tag{106}
$$

Since the above lower bound holds for all $n$, the following lower bound also holds.

$$
\liminf_{\rho \to 0} \frac{C(\rho, \beta)}{\rho^2} \ge \lim_{n \to \infty} \lim_{\rho \to 0} \frac{L_n(\rho, \beta)}{\rho^2}
\tag{107}
$$

It can be shown (see Appendix V) that

$$
\lim_{n \to \infty} \lambda_n = \lambda_\infty,
\tag{108}
$$

where $\lambda_\infty$ is given by (28). It follows from (106), (107) and (108) that

$$
\liminf_{\rho \to 0} \frac{1}{\rho^2} C(\rho, \beta) \ge \begin{cases} \frac{\lambda_\infty^2}{8} & \text{if } \frac{\lambda_\infty}{2} \le \frac{1}{\beta} \\ \left( \frac{\lambda_\infty}{2\beta} - \frac{1}{2\beta^2} \right) & \text{if } \frac{\lambda_\infty}{2} \ge \frac{1}{\beta} \end{cases}
\tag{109}
$$

## Appendix V

### Proving $\lambda_n \to \lambda_\infty$

It is easy to see that $\lambda_n \leq \lambda_\infty$ for any $n$. So,

$$\limsup_{n \to \infty} \lambda_n \geq \lambda_\infty \tag{110}$$

By assumption, $\lambda_\infty$ exists and is finite. So, for $\epsilon > 0$, there exists $N \in \mathbb{N}$ large enough such

$$\sum_{|i| < N} |R_H(i)|^2 > \lambda_\infty - \frac{\epsilon}{2}. \tag{111}$$

Let

$$K = \sum_{|i| < N} |i| |R_H(i)|^2 \tag{112}$$

For any $n > N$ such that $\frac{K}{n} < \frac{\epsilon}{2}$,

$$\lambda_n = \sum_{|i| < n} |R_H(i)|^2 (1 - \frac{|i|}{n}) \tag{113}$$

$$= \sum_{|i| < n} |R_H(i)|^2 - \frac{K}{n} - \sum_{N \leq |i| < n} |R_H(i)|^2 \frac{|i|}{n} \tag{114}$$

$$\geq \sum_{|i| < n} |R_H(i)|^2 - \frac{K}{n} - \sum_{N \leq |i| < n} |R_H(i)|^2 \tag{115}$$

$$= \sum_{|i| < N} |R_H(i)|^2 - \frac{K}{n} \tag{116}$$

$$> \lambda_\infty - \frac{\epsilon}{2} - \frac{K}{n} \tag{117}$$

$$> \lambda_\infty - \epsilon \tag{118}$$

Since $\epsilon$ is arbitrary, it follows that

$$\liminf_{n \to \infty} \lambda_n \geq \lambda_\infty \tag{119}$$

From (110) and (119), it follows that

$$\lim_{n \to \infty} \lambda_n = \lambda_\infty \tag{120}$$